\documentclass[a4paper,12pt]{article}
\usepackage{amsmath}
\usepackage{latexsym}
\newcommand{\be}{\begin{eqnarray}}
\newcommand{\ee}{\end{eqnarray}}
\newcommand{\bdm}{\begin{displaymath}}
\newcommand{\edm}{\end{displaymath}}
\begin{document}
\numberwithin{equation}{subsection}
\title{\Large\textbf{Quantum Cosmology for the General Bianchi Type
II, VI(Class A) and VII(Class A) vacuum geometries}}
\author{\textbf{T. Christodoulakis}\thanks{e-mail:
tchris@cc.uoa.gr}~~\textbf{\& ~G. O. Papadopoulos}\thanks{e-mail:
gopapado@phys.uoa.gr}}
\date{}
\maketitle
\begin{center}
\textit{University of Athens, Physics Department\\
Nuclear \& Particle Physics Section\\
Panepistimioupolis, Ilisia GR 157--71, Athens, Hellas}
\end{center}
\vspace{1cm} \numberwithin{equation}{section}
\begin{abstract}
The canonical quantization of the most general minisuperspace
actions --i.e. with all six scale factor as well as the lapse
function and the shift vector present-- describing the vacuum type
II, VI and VII geometries, is considered. The reduction to the
corresponding physical degrees of freedom is achieved through the
usage of the linear constraints as well as the quantum version of
the entire set of classical integrals of motion.
\end{abstract}
\newpage
\section{\it{Introduction}}
Since the conception by Einstein of General Relativity Theory, a
great many efforts have been devoted by many scientists to the
construction of a consistent quantum theory of gravity. These
efforts can de divided into two main approaches:
\begin{itemize}
\item[(a)] perturbative, in which one splits the metric into a background
(kinematical) part and a dynamical one:
$g_{\mu\nu}=\eta_{\mu\nu}+h_{\mu\nu}$ and tries to quantize $h_{\mu\nu}$.
The only conclusive results existing, are
that the theory thus obtained is highly nonrenormalizable \cite{goroff}.
\item[(b)] non perturbative, in which one tries to keep the twofold role
of the metric (kinematical and dynamical) intact.
A hallmark in this direction is canonical quantization.
\end{itemize}
In trying to implement this scheme for gravity, one faces the problem
of quantizing a constrained system.
The main steps one has to follow are:
\begin{itemize}
\item[(i)] define the basic operators $\widehat{g}_{\mu\nu}$
and $\widehat{\pi}^{\mu\nu}$ and the canonical
commutation relation they satisfy.
\item[(ii)] define quantum operators $\widehat{H}_{\mu}$ whose
classical counterparts are the constraint functions
$H_{\mu}$.
\item[(iii)] define the quantum states $\Psi[g]$ as the common null
eigenvector of $\widehat{H}_{\mu}$, i.e. these
satisfying $\widehat{H}_{\mu}\Psi[g]=0$. (As a consequence, one
has to check that $\widehat{H}_{\mu}$, form a
closed algebra under the basic CCR.)
\item[(iv)] find the states and define the inner product in the space of
these states.
\end{itemize}
It is fair to say that the full program has not yet been carried out,
although partial steps have been made \cite{zanelli}.

In the absence of a full solution to the problem, people have turned to
what is generally known as quantum cosmology.
This is an approximation to quantum gravity in which one freezes out all but a finite number of degrees of freedom,
and quantizes the rest. In this way one is left with a much more manageable problem that is essentially quantum mechanics
with constraints. Over the years, many models have appeared in the
literature \cite{halliwell}. In most of them,
the minisuperspace is flat and the gravitational field is represented by
no more than three degrees of freedom
(generically the three scale factors of some anisotropic Bianchi Type
model \cite{amsterdamski}).

In order for the article to be as self consistent as possible, we include in section 2, a short introduction to the theory
of constrained systems and in section 3, the Kantowski-Sachs model as an interdisciplinary example. In section 4, we
present the quantization of the most general Type II, VI(Class A)
\& VII(Class A) Vacuum Bianchi Cosmologies.
\section{\it{Elements of Constrained Dynamics}}
\subsection{\it{Introduction}}
In these short notes, we present the elements of the general
methods and some techniques of the Constrained Dynamics. It is
about a powerful mathematical theory (a method, more or less)
--primarily developed by P. A. M. Dirac. The scope of it, is to
describe singular (the definition is to be presented at the next
section) physical systems, using a generalization of the
Hamiltonian or the Lagrangian formalism. This theory, is
applicable both for discrete (i.e. finite degrees of freedom) and
continua (i.e. infinite degrees of freedom) systems.

For the sake of simplicity, the Hamiltonian point view of a physical
system is adopted, and the discussion will be restricted on discrete systems.
A basic bibliography, at which the interested reader is strongly
suggested to consult,  is quoted at the end of these notes.
Also, the treatment follows reference \cite{dirac}.
\subsection{\it{The Hamiltonian Approach}}
Suppose a discrete physical system, whose action integral is: \be
\label{action} \mathcal{A}=\int{L}dt \ee The dynamical
coordinates, are denoted by $q^{i}$, with $i \in [1,\ldots,N]$ The
Lagrangian is a function of the
coordinates and the velocities, i.e. $L=L(q^{i},\dot{q}^{i})$.\\
A note is pertinent at this point. If one demands the action
integral (\ref{action}), to be scalar under general coordinate
transformations (G.C.T.), then he can be sure that the content of
the theory to be deduced, will be relativistically covariant even
though the form of the deduced equations will not be manifestly
covariant, on account of the appearance of one particular time in
a dominant place in the theory (i.e. the time variable $t$
occurring already, as soon as one introduces the generalized
velocities, in consequently the Lagrangian, and finally the
Lagrange
transformation, in order to pass from the Lagrangian, to the Hamiltonian).\\
Variation of the action integral, gives the Euler-Lagrange equations
of motion:
\be \label{eulerlagrange}
\frac{d}{dt}\left(\frac{\partial L}{\partial \dot{q}^{i}}\right)=\frac{\partial L}{\partial q^{i}},~~~i \in [1,\ldots,N]
\ee

In order to go over to the Hamiltonian formalism, the momentum
variables $p_{i}$, are introduced through: \be
p_{i}=\frac{\partial L}{\partial \dot{q}^{i}},~~~\forall~~~ i \ee
In the usual dynamical theories, a very restricting assumption is
made; that all momenta are independent functions of the
velocities, or --in view of the inverse map theorem for a function
of many variables-- that the following (Hessian) determinant: \be
|H_{ij}|=|\frac{\partial^{2}L}{\partial \dot{q}^{i}\dot{q}^{j}}|
\ee is not zero in the whole domain of its definition. If this is
the case, then the theorem guarantees the validity of the
assumption, permits to use the Legendre transformation, and the
corresponding physical system is called \emph{Regular}. If this is
not the case (i.e. some momenta, are not independent functions od
the velocities), then there must exist some (say $M$) independent
relations of the type: \be \label{primaryconstraints}
\phi_{m}(q,p)=0,~~~m \in [1,\ldots,M] \ee which are called
\emph{Primary Constraints}. The corresponding physical systems,
are characterized as \emph{Singular}.

Variation of the quantity $p_{i}\dot{q}^{i}-L$ (the Einstein's summation
convention is in use), results in:
\be
\delta\left(p_{i}\dot{q}^{i}-L\right)=\ldots=\left(\delta p_{i}\right)\dot{q}^{i}-\left(\frac{\partial L}{\partial q^{i}}\right)
\delta q^{i}
\ee
by virtue of (\ref{eulerlagrange}). One can see that this variation,
involves variations of the $q$'s and the $p$'s. So, the
quantity under discussion does not involve variation of the velocities
and thus can be expressed in terms of the
$q$'s and the $p$'s, only. This is the Hamiltonian. It must be laid stress
on the fact that the variations, must respect the
restrictions (\ref{primaryconstraints}), i.e. to preserve them --if they are considered as conditions (see, e.g.
C. Carath\'{e}odory, `'Calculus of Variations and Partial Differential
Equations of the First Order`',
AMS Chelsea  (1989)).

Obviously, the Hamiltonian is not uniquely determined for, zero quantities can be
added to it. This means that the following:
\be \label{totalhamiltonian}
H_{T}=H+u^{m}\phi_{m}
\ee
where $u^{m}$'s are arbitrary coefficients in the phase space (including the time variable), is a valid Hamiltonian too.
Variation of (\ref{totalhamiltonian}) results in:
\be \label{equationsofmotion}
\begin{array}{l}
\dot{q}^{i}=\frac{\partial H}{\partial p_{i}}+u^{m}\frac{\partial \phi}{\partial p_{i}}+\textrm{term that vanishes as (\ref{primaryconstraints})}\\
\dot{p}_{i}=-\frac{\partial H}{\partial q_{i}}-u^{m}\frac{\partial \phi}{\partial q_{i}}-\textrm{term that vanishes as (\ref{primaryconstraints})}
\end{array}
\ee
These are the Hamiltonian equations of motion for the system under consideration. This scheme, reflects the previous
observation about variations under which, conditions must be preserved.

In order to proceed, a generalization of the Poisson Brackets is needed to be introduced. This is done as follows:\\
If $f$, $g$, $h$ are quantities on a space, endowed with a linear map $\{~,~\}$ such that:
\be \label{algebra}
\begin{array}{llll}
\{f,g\}+\{g,f\}=0 & \textrm{Antisymmetry}\\
\{f+g,h\}=\{f,h\}+\{g,h\} &  \textrm{Linearity}\\
\{fg,h\}=f\{g,h\}+\{f,h\}g &  \textrm{Product Law}\\
\{f,\{g,h\}\}+\{g,\{h,f\}\}+\{h,\{f,g\}\}=0 & \textrm{Jacobi Identity}
\end{array}
\ee
If the space is the phase space, then these Generalized Poisson Brackets, are reduced to the usual ones:
\be \label{definition}
\{f,g\}=\frac{\partial f}{\partial q^{i}}\frac{\partial g}{\partial p_{i}}-\frac{\partial g}{\partial q^{i}}\frac{\partial f}{\partial p_{i}}
\ee
otherwise are subject to the previous algebra --only.

For a dynamical variable --say $g$, one can find --with the usage of:
\be
\dot{g}=\frac{\partial g}{\partial q^{i}}\dot{q}^{i}+\frac{\partial g}{\partial p_{i}}\dot{p}_{i}
\ee
and of (\ref{equationsofmotion}), as well as the generalized Poisson Bracket Algebra (\ref{algebra}):
\be  \label{motionofg}
\dot{g}\approx \{g,H_{T}\}
\ee
The symbol $\approx$ is the \emph{Weak Equality} symbol and stands for the following rule (deduced from the
thorough analysis of the previous procedure):\\
\emph{A constraint, must not be used before all the Generalized Poisson Brackets, are calculated formally (i.e. only
with the usage of the algebra (\ref{algebra}) and the usual definition (\ref{definition}) --when the last is applicable)}.
This rule, is encoded as:
\be
\phi_{m}(q,p)\approx 0,~~~m \in [1,\ldots,M]
\ee
In the previous procedure, the position of that rule, reflects the need to manipulate the $u^{m}$'s, which may depend
on $t$ only --since they are unknown coefficients, the definition (\ref{definition}) can not be used.

If the dynamical variable $g$ is any one of the constraints, then (\ref{primaryconstraints}) declare the preservation
of zero. Thus, consistency conditions, are deduced:
\be \label{consistencyconditions}
\{\phi_{m'},H\}+u^{m}\{\phi_{m'},\phi_{m}\}=0
\ee
There are three possibilities:
\begin{itemize}
\item[$CC_{1}$] Relations (\ref{consistencyconditions}) lead to identities --maybe, with the help of
(\ref{primaryconstraints}).
\item[$CC_{2}$] Relations (\ref{consistencyconditions}) lead to equations independent of the $u$'s. These turn
to be constraints also. They called \emph{Secondary}, but must be treated on the same footing as the primary ones.
\item[$CC_{3}$] Relations (\ref{consistencyconditions}) impose conditions on the $u$'s.
\end{itemize}
The above procedure must be applied for all the secondary constraints. Again, the possible cases will be the previous
three. The new constraints which may turn up are called secondary too.  The procedure is applied for once more and so on.
At the end, one will have a number of constraints (primary plus secondary) --say $\mathcal{J}$-- and a number of conditions on the $u$'s.
A detailed analysis of the set of these conditions, shows that:
\be \label{us}
u^{m}=U^{m}(q,p)+\mathcal{V}^{a}(t)V^{m}_{a}(q,p)
\ee
where $V^{m}_{a}(q,p)$ are the $a$ (in number) independent solutions of the homogeneous systems:
\bdm
V^{m}_{a}(q,p)\{\phi_{m''},\phi_{m}\}=0
\edm
The functions $\mathcal{V}^{a}(t)$ are related to the gauge freedom of the physical system.

Some terminology is needed at this point.\\
A dynamical variable $R$, is said to be \emph{First Class}, if it has zero Poisson Bracket, with \underline{all} the
constraints:
\be
\{R,\phi_{n}\}=0,~~~n \in [1,\ldots,\mathcal{J}]
\ee
where $\mathcal{J}$ is the total number of constraints --i.e. primary plus all the secondary ones.
It is sufficient for these conditions, to hold weakly --since, by definition, the $\phi$'s are the only independent
quantities that vanish weakly. Otherwise, the variable $R$, is said to be \emph{Second Class}.
If  $R$, is First Class, then the quantity $\{R,\phi_{n}\}$ is strongly equal to some linear combination of the $\phi$'s.
The following relative theorem (with a trivial proof) holds:\\
\emph{''The Poisson Bracket of two First Class quantities, is also First Class''.}

Using the result (\ref{us}) the Hamiltonian
(\ref{totalhamiltonian}), which is called \emph{Total
Hamiltonian}, is written: \be
H_{T}=H+U^{m}\phi_{m}+\mathcal{V}^{a}V^{m}_{a}\phi_{m}\equiv H'
+\mathcal{V}^{a}\phi_{a} \ee with obvious associations. It can be
proved that $H'$ and $\phi_{a}$, are first class quantities.
With this splitting and the relation (\ref{motionofg}) for a dynamical variable $g$, it can be deduced that:\\
\emph{The First Class Primary Constraints $\phi_{a}$, are the
generating functions (i.e. the quantities $\{g,\phi_{a}\}$) of
infinitesimal Contact Transformations; i.e. of transformations
which lead to changes in the $q$'s and the $p$'s, but they do not
affect the physical state of the system}.

Successive application of two contact transformations generated by
two given First Class Primary Constraints and taking into account
the order, leads --for the sake of consistency-- to a new
generating function: $\{g,\{\phi_{a},\phi_{a'}\}\}$. Thus one can
see that First Class Secondary Constraints, which may turn up from
$\{\phi_{a},\phi_{a'}\}$, can also serve as generating functions
of infinitesimal Contact Transformations. Possibly, another way to
produce  First Class Secondary Constraints, is the First Class
quantity $\{H',\phi_{a}\}$. Since no one has found an example of a
First Class Secondary Constraint, which affects the physical state
when used as generating function, the conclusion is that all First
Class quantities, are  generating functions of infinitesimal
Contact Transformations. Thus, the total Hamiltonian should be
replaced by the \emph{Extended Hamiltonian} $H_{E}$, defined as:
\be H_{E}=H_{T}+\mathcal{U}^{a''}\phi_{a''} \ee where the
$\phi_{a''}$'s are those First Class Secondary Constraints, which
are not already included in $H_{T}$. Finally, the equation of
motion for a dynamical variable $g$ (\ref{motionofg}) is altered:
\be \dot{g}\approx\{g,H_{E}\} \ee
\subsection{\it{Quantization of Constrained Systems}}
\subsubsection{\it{No Second Class Constraints are Present}}
The quantization of a classical physical system, whose Lagrangian, gives first class constraints only, is made in three steps:
\begin{itemize}
\item[$S_{1}$] The dynamical coordinates $q$'s and momenta $p$'s, are turned into Hermitian Operators
$\widehat{q}$~'s and $\widehat{p}$~'s, satisfying
the basic commutative algebra: $[\widehat{q}^{i},\widehat{p}_{j}]=i\delta^{i}_{j}$.
\item[$S_{2}$] A kind of a  Schr\"{o}dinger equation, is set up.
\item[$S_{3}$] Any dynamical function, become Hermitian Operator --provided that the ordering problem is
somehow solved.
\end{itemize}
Obviously, the constraints --being functions on the phase space-- are subject to the $S_{3}$ rule. Dirac, proposed
that when the constraints are turned into operators, they must annihilate the wave function $\Psi$:
\be \label{annihilation}
\widehat{\phi}_{i}\Psi=0,~~~\forall~~~i
\ee
Successive application of two such  given conditions and taking into account the order, for sake of consistency, results in:
\be \label{quantumconditions}
[\widehat{\phi}_{i},\widehat{\phi}_{j}]\Psi=0
\ee
In order for operational conditions (\ref{quantumconditions}) not to give new ones on $\Psi$, one demands:
\be
[\widehat{\phi}_{i},\widehat{\phi}_{j}]=C^{k}_{ij},\widehat{\phi}_{k}
\ee
If it is possible for such an algebra to be deduced, then no new operational conditions on $\Psi$ are found and the system
is consistent. If this is not the case, the new conditions must be taken into account and along with the initial ones,
must give closed algebra, otherwise the procedure must be continued until a closed algebra is found.
The discussion  does not end here. Consistency between the operational conditions (\ref{annihilation}) and the
Schr\"{o}dinger equation, is pertinent as well. This lead to:
\be
[\widehat{\phi}_{i},\widehat{H}]\Psi=0
\ee
and consistency know, reads:
\be
[\widehat{\phi}_{i},\widehat{H}]=D^{k}_{i},\widehat{\phi}_{k}
\ee
\subsubsection{\it{Second Class Constraints are Present}}
Suppose we have a classical physical system, whose Lagrangian,
gives second class constraints. Any set of constraints, can be
replaced by a corresponding set of independent linear combinations
of them. It is thus, in principle, possible to make arrangement
such that the final set of constraints, contains as much first
class constraints as possible. Using the remaining --say $S$ in
number-- second class constraints, the following matrix is
defined: \be \Delta_{ij}=[\chi_{i},\chi_{j}],~~~(i,j) \in
[1,\ldots,S] \ee
where $,\chi$'s are the remaining (in classical form)  second class constraints. A theorem can be proved :\\
\emph{''The determinant of this matrix does not vanish, not even weakly''.}

Since the determinant of $\Delta$ is non zero, there is the inverse of this matrix; say $\Delta^{-1}$.
Dirac, proposed a new kind of Poisson Bracket, the $\{~,~\}_{D}$:
\be
\{~,~\}_{D}=\{~,~\}-\sum^{S}_{i=1}\sum^{S}_{j=1}\{~,\chi_{i}\}\Delta^{-1}_{ij}\{\chi_{j},~\}
\ee
These Brackets, are antisymmetric, linear in their arguments, obey the product law and the Jacobi identity.
It holds that:
\be
\{g,H_{E}\}_{D}\approx \{g,H_{E}\}
\ee
because terms like $\{\chi_{i},H_{E}\}$, with $H_{E}$ being first class, vanish weakly.
Thus:
\be
\dot{g}\approx\{g,H_{E}\}_{D}
\ee
But:
\be
\{\xi,\chi_{s}\}_{D}=\ldots=0
\ee
if $\xi$ is any of the $q$'s or the $p$'s. Thus, at the classical level, one may put the second class constraints equal to
zero, before calculating the new Poisson Brackets. That means that:
\begin{itemize}
\item[$M_{1}$] The equations $\chi=0$ may be considered as strong equations.
\item[$M_{2}$] One, must ignore the corresponding degrees of freedom and
\item[$M_{3}$] quantize the rest, according to the general rules, given in the previous section.
\end{itemize}
\section{An Interdisciplinary Example:\\
The Kantowski-Sachs Model} The purpose of the present section is
to illustrate an application of the Dirac's method for constrained
systems. The example chosen, is that of Kantowski-Sachs reduced
Lagrangian --i.e. of a vacuum cosmological model; thus the
interdisciplinary character of the section, emerges.

Consider, the following Lagrangian: \be
L=-\frac{a(t)\dot{b}^{2}(t)+2b(t)\dot{a}(t)\dot{b}(t)}{2N(t)}+\frac{N(t)a(t)}{2}
\ee where $a(t)$, $b(t)$ and $N(t)$, are the three degrees of
freedom (the $q$'s). It must be brought to the reader's notice
that the Euler-Lagrange equations corresponding to this
Lagrangian, are tantamount to the Einstein's Field Equations for
the Kantowski-Sachs model (described in \cite{kantowski}),
characterized by the line element: \be
ds^{2}=-N^{2}(t)dt^{2}+a^{2}(t)dr^{2}+b^{2}(t)d\theta^{2}+b^{2}(t)sin^{2}(\theta)d\phi^{2}
\ee

The momenta are:
\be \label{momenta}
p_{a}=\frac{\partial L}{\partial \dot{a}(t)} & =  &-\frac{b(t)\dot{b}(t)}{N(t)}\\
p_{b}=\frac{\partial L}{\partial \dot{b}(t)} &  =  & -\frac{\dot{a}(t)b(t)+a(t)\dot{b}(t)}{N(t)}\\
p_{N}= \frac{\partial L}{\partial \dot{N}(t)} &  = & 0
\ee
>From the third of (\ref{momenta}), one can see that there is one primary constraint:
\be
p_{N}\approx 0
\ee

The total Hamiltonian is:
\be
H_{T}=H+u(a(t),b(t),N(t),t)p_{N}
\ee
where:
\be
H=p_{a}\dot{a}(t)+p_{b}\dot{b}(t)-L=N(t)\Omega(t)
\ee
with:
\be \label{omega}
\Omega(t)\equiv-\frac{a(t)}{2}-\frac{p_{a}p_{b}}{b(t)}+\frac{a(t)p^{2}_{a}}{2b^{2}(t)}
\ee

The consistency condition (\ref{consistencyconditions}) applied to:
\begin{itemize}
\item[$A_{1}$] the constraint $p_{N}\approx 0$, gives one secondary constraint:
\be \label{chiconstraint}
\chi\equiv\{p_{N},H\}=\{p_{N},N(t)\Omega(t)\}=-\Omega(t)\approx 0
\ee
A straightforward calculation, results in:
\be \label{chipn}
\{\chi,p_{N}\}=0
\ee
\item[$A_{2}$] the previously deduced secondary constraint $\chi\approx 0$, gives --by virtue of (\ref{omega}),
(\ref{chiconstraint}) and (\ref{chipn})--  no further constraints, since it is identically satisfied ($CC_{1}$ case):
\be
\{\chi,H\}+u(a(t),b(t),N(t),t)\{\chi,p_{N}\}=0
\ee
\end{itemize}

The Poisson Bracket (\ref{chipn}) also declares that both $p_{N}$ and $\chi$, are first class quantities.

Finally, the equations of motion are:
\be
\dot{a}(t)\approx \{a(t),H_{T}\}
\ee
\be
\dot{p_{a}}\approx \{p_{a},H_{T}\}
\ee
\be
\dot{b}(t)\approx \{b(t),H_{T}\}
\ee
\be
\dot{p_{b}}\approx \{p_{b},H_{T}\}
\ee
\be \label{multiplier}
\dot{N}(t)\approx \{N(t),H_{T}\}
\ee
\be \label{pn}
\dot{p_{N}}\approx \{p_{N},H_{T}\}
\ee
The first four equations constitute the usual set of the Euler-Lagrange equations for the $a(t)$ and $b(t)$, degrees of
freedom. Equation (\ref{multiplier}), results in the gauge freedom related to $N(t)$ since --according to this equation--
$\dot{N}(t)=u(a(t),b(t),N(t),t)$, i.e. an arbitrary function of time, while equation (\ref{pn}) is trivially satisfied, in view
of (\ref{chiconstraint}).
\newpage
\section{\it{Quantization of the most general Bianchi Type II,
\\ VI(Class A) \& VII(Class A) Vacuum Cosmologies}}

We firstly treat of (see T. Christodoulakis, G. O. Papadopoulos,
Phys. Lett. B \textbf{501} (2001) 264-8):
\subsection{\it{The Bianchi Type II Case }}
In \cite{tchristype2}, we had considered the quantization of an action corresponding to the most general
Bianchi Type II cosmology, i.e. an action giving Einstein's Field Equations, derived from the line
element:
\be
ds^{2}=(N^{2}(t)-N_{a}(t)N^{a}(t))dt^{2}+2N_{a}(t)\sigma^{a}_{i}(x)dx^{i}dt+
\gamma_{\alpha\beta}(t)\sigma^{\alpha}_{i}(x)\sigma^{\beta}_{j}(x)dx^{i}dx^{j}
\ee
with:
\be
\begin{array}{l}
  \sigma^{a}(x)=\sigma^{\alpha}_{i}(x)dx^{i}\\
  \sigma^{1}(x)=dx^{2}-x^{1}dx^{3} \\
  \sigma^{2}(x)=dx^{3} \\
  \sigma^{3}(x)=dx^{1} \\
  d\sigma^{a}(x)=\frac{1}{2}C^{a}_{\beta\gamma}\sigma^{\beta}\wedge\sigma^{\gamma}\\
  C^{1}_{23}=-C^{1}_{32}=1
\end{array}
\ee
see \cite{ryan}.

As is well known \cite{sneddon}, the Hamiltonian is
$H=\widetilde{N}(t)H_{0}+N^{a}(t)H_{a}$ where:
\be \label{6paper3}
H_{0}=\frac{1}{2}L_{\alpha\beta\mu\nu}\pi^{\alpha\beta}\pi^{\mu\nu}+\gamma
R
\ee
is the quadratic constraint with:
\be
\begin{array}{l}
  L_{\alpha\beta\mu\nu}=\gamma_{\alpha\mu}\gamma_{\beta\nu}+\gamma_{\alpha\nu}\gamma_{\beta\mu}-
\gamma_{\alpha\beta}\gamma_{\mu\nu} \\
  R=C^{\beta}_{\lambda\mu}C^{\alpha}_{\theta\tau}\gamma_{\alpha\beta}\gamma^{\theta\lambda}
\gamma^{\tau\mu}+2C^{\alpha}_{\beta\delta}C^{\delta}_{\nu\alpha}\gamma^{\beta\nu}+
4C^{\mu}_{\mu\nu}C^{\beta}_{\beta\lambda}\gamma^{\nu\lambda}=
C^{\alpha}_{\mu\kappa}C^{\beta}_{\nu\lambda}\gamma_{\alpha\beta}\gamma^{\mu\nu}\gamma^{\kappa\lambda}
\end{array}
\ee
$\gamma$ being the determinant of $\gamma_{\alpha\beta}$ (the last
equality holding only for the Type II case), and:
\be
H_{a}=C^{\mu}_{a\rho}\gamma_{\beta\mu}\pi^{\beta\rho}
\ee
are the linear constraints. Note that $\widetilde{N}$ appearing in
the Hamiltonian, is to be identified with $N/\sqrt{\gamma}$.

The quantities $H_{0}$, $H_{a}$, are weakly vanishing \cite{dirac},
i.e. $H_{0}\approx 0$, $H_{a}\approx 0$. For all class A Bianchi
Types ($C^{\alpha}_{\alpha\beta}=0$), it can be seen to obey the
following first-class algebra:
\be
\begin{array}{l}
  \{H_{0}, H_{0}\}=0 \\
  \{H_{0}, H_{a}\}=0 \\
  \{H_{a}, H_{\beta}\}=-\frac{1}{2}C^{\gamma}_{\alpha\beta}H_{\gamma}
\end{array}
\ee
which ensures their preservation in time i.e. $\dot{H}_{0}\approx
0$, $\dot{H}_{a}\approx 0$ and establishes the consistency of the
action.

If we follow Dirac's general proposal \cite{dirac} for quantizing
this action, we have to turn $H_{0}$, $H_{a}$, into operators
annihilating the wave function $\Psi$.

In the Schr\"{o}dinger representation:
\be
\begin{array}{l}
  \gamma_{\alpha\beta}\rightarrow
\widehat{\gamma}_{\alpha\beta}=\gamma_{\alpha\beta} \\
  \pi^{\alpha\beta}\rightarrow
\widehat{\pi}^{\alpha\beta}=-i\frac{\partial}{\partial\gamma_{\alpha\beta}}
\end{array}
\ee
satisfying the basic Canonical Commutation Relation (CCR)
--corresponding to the classical ones:
\be
[\widehat{\gamma}_{\alpha\beta},
\widehat{\pi}^{\mu\nu}]=-i\delta^{\mu\nu}_{\alpha\beta}=\frac{-i}{2}
(\delta^{\mu}_{\alpha}\delta^{\nu}_{\beta}+\delta^{\mu}_{\beta}\delta^{\nu}_{\alpha})
\ee

The quantum version of the 2 independent linear constraints has
been used to reduce, via the method of characteristics \cite{carabedian},
the dimension of the initial configuration space from 6
($\gamma_{\alpha\beta}$) to 4 (combinations of
$\gamma_{\alpha\beta}$), i.e.
$\Psi=\Psi(q,\gamma,\gamma^{2}_{12}-\gamma_{11}\gamma_{22},\gamma_{12}\gamma_{13}-
\gamma_{11}\gamma_{23})$, where
$q=C^{\alpha}_{\mu\kappa}C^{\beta}_{\nu\lambda}\gamma_{\alpha\beta}\gamma^{\mu\nu}\gamma^{\kappa\lambda}$.

According to Kuha\v{r}'s and Hajicek's \cite{hajiceck} prescription,
the `'kinetic`' part of $H_{0}$ is to be realized as the
conformal Laplacian, corresponding to the reduced metric:
\begin{equation}
L_{\alpha\beta\mu\nu}\frac{\partial x^{i} }{\partial
\gamma_{\alpha\beta} }\frac{\partial x^{j}}{\partial
\gamma_{\mu\nu}}=g^{ij}
\end{equation}
where $x^{i}$, $i=1,2,3,4$, are the arguments of $\Psi$. The
solutions had been presented in \cite{tchristype2}. Note that the
first-class algebra satisfied by $H_{0}$, $H_{a}$, ensures that
indeed, all components of $g^{ij}$ are functions of the $x^{i}$'s.
The signature of the $g^{ij}$, is $(+, +, -, -)$ signaling the
existence of gauge degrees of freedom among the $x^{i}$'s.

Indeed, one can prove \cite{tchriskuchar} that the only gauge invariant
quantity which, uniquely and irreducibly, characterizes a
3-dimensional geometry admitting the Type II symmetry group, is:
\be
q=C^{\alpha}_{\mu\kappa}C^{\beta}_{\nu\lambda}\gamma_{\alpha\beta}\gamma^{\mu\nu}\gamma^{\kappa\lambda}
\ee
An outline of the proof, is as follows:\\
Let two hexads $\gamma^{(1)}_{\alpha\beta}$ and
$\gamma^{(2)}_{\alpha\beta}$ be given, such that their
corresponding $q$'s, are equal. Then \cite{tchriskuchar} there exists an
automorphism matrix $\Lambda$ (i.e. satisfying
$C^{a}_{\mu\nu}\Lambda^{\kappa}_{a}=C^{\kappa}_{\rho\sigma}\Lambda^{\rho}_{\mu}\Lambda^{\sigma}_{\nu}$)
connecting them, i.e.
$\gamma^{(1)}_{\alpha\beta}=\Lambda^{\mu}_{\alpha}\gamma^{(2)}_{\mu\nu}\Lambda^{\nu}_{\beta}$.
But as it had been shown in the appendix of \cite{tchristype5}, this kind of
changes on $\gamma_{\alpha\beta}$, can be seen to be induced by
spatial diffeomorphisms. Thus, 3-dimensional Type II geometry, is
uniquely characterized by some value of $q$.

Although for full pure gravity, Kuchar\v{r} \cite{kuchar} has shown that
there are not other first-class functions, homogeneous and linear
in $\pi^{\alpha\beta}$, except $H_{a}$, imposing the extra
symmetries (Type II), allows for such quantities to exist --as
it will be shown. We are therefore, naturally led to seek the
generators of these extra symmetries --which are expected to chop
off $x^{2}$, $x^{3}$, $x^{4}$. Such quantities are, generally,
called in the literature `'Conditional Symmetries`'.

The automorphism group for Type II, is described by the following
6 generators --in matrix notation and collective form:
\be
\lambda^{a}_{(I)\beta}=\left(
\begin{array}{ccc}
  \kappa+\mu & x & y \\
  0 & \kappa & \rho \\
  0 & \sigma & \mu
\end{array}\right)
\ee
with the property:
\be
C^{a}_{\mu\nu}\lambda^{\kappa}_{a}=C^{\kappa}_{\mu\sigma}\lambda^{\sigma}_{\nu}+C^{\kappa}_{\sigma\nu}\lambda^{\sigma}_{\mu}
\ee
>From these matrices, we can construct the linear --in momenta--
quantities:
\be
A_{(I)}=\lambda^{a}_{(I)\beta}\gamma_{\alpha\rho}\pi^{\rho\beta}
\ee
Two of these, are the $H_{a}$,'s since $C^{a}_{(\rho)\beta}$
correspond to the inner automorphism subgroup --designated by the
x and y parameters, in $\lambda^{a}_{(I)\beta}$. The rest of
them, are the generators of the outer automorphisms and are
described by the matrices:
\be
\varepsilon^{a}_{(I)\beta}=\left(\begin{array}{ccc}
  \kappa+\mu & 0 & 0 \\
  0 & \kappa & \rho \\
  0 & \sigma & \mu
\end{array}\right)
\ee
The corresponding --linear in momenta-- quantities, are:
\be
E_{(I)}=\varepsilon^{a}_{(I)\beta}\gamma_{\alpha\rho}\pi^{\rho\beta}
\ee
The algebra of these --seen as functions on the phase space,
spanned by $\gamma_{\alpha\beta}$ and $\pi^{\mu\nu}$--, is:
\be \label{15paper2}
\begin{array}{l}
  \{E_{I}, E_{J}\}=\widetilde{C}^{K}_{IJ}E_{K} \\
  \{E_{I}, H_{a}\}=-\frac{1}{2}\lambda^{\beta}_{a}H_{\beta} \\
  \{E_{I}, H_{0}\}=-2(\kappa+\mu)\gamma R
\end{array}
\ee
>From the last of  (\ref{15paper2}), we conclude that the subgroup of $E_{I}$'s
with the property $\kappa+\mu=0$, i.e. the traceless generators,
are first-class quantities; their time derivative vanishes. So
let:
\be
\widetilde{E}_{I}=\{E_{I}:~\kappa+\mu=0\}
\ee
Then, the previous statement translates into the form:
\be \label{17paper2}
\dot{\widetilde{E}_{I}}=0 \Rightarrow \widetilde{E}_{I}=c_{I}
\ee
the $c_{I}$'s being arbitrary constants.

Now, these are --in principle-- integrals of motion. Since, as we
have earlier seen, $\widetilde{E}_{I}$'s along with $H_{a}$'s,
generate automorphisms, it is natural to promote the integrals of
motion (\ref{17paper2}), to symmetries --by setting the $c_{I}$'s zero. The
action of the quantum version of these $\widetilde{E}_{I}$'s on
$\Psi$, is taken to be \cite{hajiceck}:
\be
\begin{array}{l}
  \widehat{\widetilde{E}}_{I}\Psi=\varepsilon^{a}_{(I)\beta}\gamma_{\alpha\rho}\frac{\partial\Psi}{\gamma_{\beta\rho}}=0 \\
  \varepsilon^{a}_{(I)a}=0
\end{array} \Bigg\}\Rightarrow \Psi=\Psi(q,\gamma)
\ee

The Wheeler-DeWitt equation now, reads:
\be \label{19paper2}
5q^{2}\frac{\partial^{2} \Psi}{\partial
q^{2}}-3\gamma^{2}\frac{\partial^{2} \Psi}{\partial
\gamma^{2}}+2q\gamma\frac{\partial^{2}\Psi}{\partial
\gamma\partial q}+5q\frac{\partial \Psi}{\partial
q}-3\gamma\frac{\partial \Psi}{\partial \gamma}-2q\gamma\Psi=0
\ee
\textit{Note that:
\be
\nabla^{2}_{c}=\nabla^{2}+\frac{(d-2)}{4(d-1)}R=\nabla^{2}
\ee
since we have a 2-dimensional, flat space, with contravariant
metric:
\be
g^{ij}=\left(\begin{array}{cc}
  5q^{2} & q\gamma \\
  q\gamma & -3\gamma^{2}
\end{array}\right)
\ee
which is Lorentzian}. This equation, can be easily solved by
separation of variables; transforming to new coordinates
$u=q\gamma^{3}$ and $v=q\gamma$, we get the 2 independent
equations:
\be
\begin{array}{l}
  16u^{2}A''(u)+16uA'(u)-cA(u)=0 \\
  B''(v)+\frac{1}{v}B'(v)-(\frac{1}{2v}+\frac{c}{4v^{2}})B(v)=0
\end{array}
\ee where c, is the separation constant. Equation
(\ref{19paper2}), is of hyperbolic type and the resulting wave
function will still not be square integrable. Besides that, the
tracefull generators of the outer automorphisms, are left inactive
--due to the non vanishing CCR with $H_{0}$.

These two facts, lead us to deduce that there must still exist a
gauge symmetry, corresponding to some --would be, linear in
momenta-- first-class quantity. Our starting point in the pursuit
of this, is the third of (\ref{15paper2}). It is clear that we need another
quantity --also linear in momenta-- with an analogous property;
the trace of $\pi^{\mu\nu}$, is such an object. We thus define
the following quantity:
\be
T=E_{I}-(\kappa+\mu)\gamma_{\alpha\beta}\pi^{\alpha\beta}
\ee
in the phase space --spanned by $\gamma_{\alpha\beta}$ and
$\pi^{\mu\nu}$. It holds that:
\be
\begin{array}{l}
  \{T, H_{0}\}=0 \\
  \{T, H_{a}\}=0 \\
  \{T, E_{I}\}=0
\end{array}
\ee
because of:
\be
\begin{array}{l}
  \{E_{I}, \gamma\}=-2(\kappa+\mu)\gamma \\
  \{E_{I}, q\}=0 \\
  \gamma_{\alpha\beta}\{\pi^{\alpha\beta}, q\}=q \\
  \gamma_{\alpha\beta}\{\pi^{\alpha\beta}, \gamma\}=-3\gamma
\end{array}
\ee

Again --as for $\widetilde{E}_{I}$'s--, we see that since $T$, is
first-class, we have that:
\be
\dot{T}=0 \Rightarrow T=const=c_{T}
\ee
another integral of motion. We therefore see, that $T$ has all
the necessary properties to be used in lieu of the tracefull
generator, as a symmetry requirement on $\Psi$. In order to do
that, we ought to set $c_{T}$ zero --exactly as we did with the
$c_{I}$'s, corresponding to $\widetilde{E}_{I}$'s. The quantum
version of $T$, is taken to be:
\be
\widehat{T}=\lambda^{\alpha}_{\beta}\gamma_{\alpha\rho}\frac{\partial}{\partial
\gamma_{\beta\rho}}-(\kappa+\mu)\gamma_{\alpha\beta}\frac{\partial}{\partial
\gamma_{\alpha\beta}}
\ee
Following, Dirac's theory, we require:
\be \label{27paper2}
\widehat{T}\Psi=\lambda^{\alpha}_{\beta}\gamma_{\alpha\rho}\frac{\partial
\Psi}{\partial
\gamma_{\beta\rho}}-(\kappa+\mu)\gamma_{\alpha\beta}\frac{\partial
\Psi}{\partial
\gamma_{\alpha\beta}}=(\kappa+\mu)(q\frac{\partial\Psi}{\partial
q}-\gamma\frac{\partial\Psi}{\partial \gamma})=0
\ee
Equation (\ref{27paper2}), implies that $\Psi(q,\gamma)=\Psi(q\gamma)$ and
thus equation (\ref{19paper2}), finally, reduces to:
\be \label{28paper2}
4w^{2}\Psi''(w)+4w\Psi'(w)-2w\Psi=0
\ee
where, for simplicity, $w\doteq q\gamma$. The solution to this
equation, is:
\be
\Psi=c_{1}I_{0}(\sqrt{2q\gamma})+c_{2}K_{0}(\sqrt{2q\gamma})
\ee
where $I_{0}$ is the modified Bessel function, of the first kind,
and $K_{0}$ is the modified Bessel function, of the second kind,
both with zero argument.

At first sight, it seems that although we have apparently
exhausted the symmetries of the system, we have not yet been able
to obtain a wave function on the space of the 3-geometries, since
$\Psi$ depends on $q\gamma$ and not on $q$ only. On the other
hand, the fact that we have achieved a reduction to one degree of
freedom, must somehow imply that the wave function found must be a
function of the geometry. This puzzle finds its resolution as
follows. Consider the quantity: \be
\Omega=-2\gamma_{\rho\sigma}\pi^{\rho\sigma}+{
\frac{2C^{a}_{\mu\kappa}C^{\beta}_{\nu\lambda}\gamma^{\kappa\lambda}\gamma^{\mu\nu}
\gamma_{\alpha\rho}\gamma_{\beta\sigma}-4C^{\alpha}_{\mu\rho}C^{\beta}_{\nu\sigma}\gamma_{\alpha\beta}\gamma^{\mu\nu}}{q}}\pi^{\rho\sigma}
\ee This can also be seen to be first-class, i.e. \be
\dot{\Omega}=0 \Rightarrow \Omega=const=c_{\Omega} \ee Moreover,
it is a linear combination of $T$, $\widetilde{E}_{I}$'s, and
$H_{a}$'s, and thus $c_{\Omega}=0$. Now it can be verified that
$\Omega$, is nothing but: \be
\frac{1}{N(t)}(\frac{\dot{\gamma}}{\gamma}+\frac{1}{3}\frac{\dot{q}}{q})
\ee So: \be \gamma q^{1/3}=\vartheta=constant \ee Without any loss
of generality, and since $\vartheta$ is not an essential constant
of the classical system (see \cite{tchrisaut} and reference [18]
therein), we set $\vartheta=1$. Therefore: \be
\Psi=c_{1}I_{0}(\sqrt{2}q^{1/3})+c_{2}K_{0}(\sqrt{2}q^{1/3}) \ee
where $I_{0}$ is the modified Bessel function, of the first kind,
and $K_{0}$ is the modified Bessel function, of the second kind,
both with zero argument.

As for the measure, it is commonly accepted that, there is not a
unique solution. A natural choice, is to adopt the measure that
makes the operator in (\ref{28paper2}), hermitian --that is:
\be
\mu(q)\propto q^{-1}
\ee
It is easy to find combinations of $c_{1}$ and $c_{2}$ so that
the probability $\mu(q)|\Psi|^{2}$, be defined.

Note that putting the constant associated with $\Omega$, equal to
zero, amounts in restricting to a subset of the classical
solutions, since $c_{\Omega}$, is one of the two essential
constants of Taub's solution. One could keep that constant, at
the expense of arriving at a wave function with explicit time
dependence, since then:
\be
\gamma=q^{-1/3}Exp[\int{c_{\Omega}N(t)}dt]
\ee
We however, consider more appropriate to set that constant zero,
thus arriving at a $\Psi$ depending on $q$ only, and decree its
applicability to the entire space of the classical solutions.
Anyway this is not such a blunder, since $\Psi$ is to give weight
to all states, --being classical ones, or not.

\newpage
The above treatise,  remains almost unchanged for --(T. Chistodoulakis, G. O. Papadopoulos, to appear in
Phys. Lett. B):
\subsection{\it{The Class A Bianchi Type VI \& VII Case}}

The only difference between the various Bianchi Types, is that of
the values of the structure constants. Of course, it must be
emphasized that in the present work we study differential
equations (i.e. the Einstein's Field Equations), which are defined
locally --we are not concerned with global properties (i.e.
topological aspects).

It is well known that in 3 dimensions, the tensor
$C^{\alpha}_{\beta\gamma}$ admits a unique decomposition in terms
of a contravariant symmetric tensor density of weight $-1$,
$m^{\alpha\beta}$, and a covariant vector
$\nu_{\alpha}=\frac{1}{2}C^{\rho}_{\alpha\rho}$ as follows:
\be
C^{\alpha}_{\beta\gamma}=m^{\alpha\delta}\varepsilon_{\delta\beta\gamma}+\nu_{\beta}\delta^{\alpha}_{\gamma}-\nu_{\gamma}\delta^{\alpha}_{\beta}
\ee
For the Class A $(\nu_{\alpha}=0)$ Bianchi Type VI
$(\varepsilon=1)$ and VII $(\varepsilon=-1)$, this matrix is:
\be
m^{\alpha\beta}=\left(\begin{array}{ccc}
  \varepsilon & 0 & 0 \\
  0 & -1 & 0 \\
  0 & 0 & 0
\end{array}\right)
\ee
resulting in the following non vanishing structure constants
(\cite{ryan}):
\be
\begin{array}{cc}
  C^{1}_{23}=\varepsilon & ~~~C^{2}_{13}=1
\end{array}
\ee

In general, via the method of characteristics \cite{carabedian}, the
quantum version of the three independent linear constraints can be
used to reduce the dimension of the initial configuration space
from 6 ($\gamma_{\alpha\beta}$) to 3 (combinations of
$\gamma_{\alpha\beta}$), i.e. $\Psi=\Psi(q^{1},q^{2},q^{3})$
\cite{hajiceck}, where:
\be \label{12paper3}
\begin{array}{l}
  q^{1}=C^{\alpha}_{\mu\kappa}C^{\beta}_{\nu\lambda}\gamma^{\mu\nu}\gamma^{\kappa\lambda}\gamma_{\alpha\beta} \\
  q^{2}=C^{\alpha}_{\beta\kappa}C^{\beta}_{\alpha\lambda}\gamma^{\kappa\lambda} \\
  q^{3}=\gamma
\end{array}
\ee In analogy with the preceding case (where an outline of the
proof is exhibited), the only G.C.T. (gauge) invariant quantities,
which uniquely and irreducibly, characterize a 3-dimensional
geometry admitting the Class A Type VI and VII symmetry groups,
are the quantities $q^{1}$ and $q^{2}$.

In terms of the three $q$~'s, one can define the following
--according to \cite{hajiceck}-- induced `'physical`' metric, given by
the relation:
\be
g^{ij}=L_{\alpha\beta\mu\nu}\frac{\partial q^{i} }{\partial
\gamma_{\alpha\beta}}\frac{\partial q^{j}}{\partial
\gamma_{\mu\nu}}=\left(\begin{array}{ccc}
  5q^{1}q^{1}-16q^{2}q^{2} & q^{1}q^{2} & q^{1}q^{3} \\
  q^{1}q^{2} & q^{2}q^{2} & q^{2}q^{3} \\
  q^{1}q^{3} & q^{2}q^{3} & -3q^{3}q^{3}
\end{array}\right)
\ee
Note that the first-class algebra satisfied by $H_{0}$,
$H_{\alpha}$, ensures that indeed, all components of $g^{ij}$ are
functions of the $q^{i}$.

The automorphism group for the Class A Type VI and VII, is
described by the following 4 generators --in matrix notation and
collective form:
\be
\lambda^{\alpha}_{(I)\beta}=\left(
\begin{array}{ccc}
  a  & \varepsilon~c & b \\
  c & a & d \\
  0 & 0 & 0
\end{array}\right)
\ee
>From these matrices, we can construct the linear --in momenta--
quantities:
\be
A_{(I)}=\lambda^{\alpha}_{(I)\beta}\gamma_{\alpha\rho}\pi^{\rho\beta}
\ee
Three of these are the $H_{\alpha}$'s, since
$C^{\alpha}_{(\rho)\beta}$ correspond to the inner automorphism
subgroup --designated by the $c$, $b$ and $d$ parameters in
$\lambda^{\alpha}_{(I)\beta}$. The remaining is the generator of
the outer automorphisms and is given by the essentially unique
matrix
\be \label{17paper3}
\epsilon^{\alpha}_{\beta}=\left(\begin{array}{ccc}
  a & 0 & 0 \\
  0 & a & 0 \\
  0 & 0 & 0
\end{array}\right)
\ee
The corresponding --linear in momenta-- quantity is:
\be
E=\epsilon^{\alpha}_{\beta}\gamma_{\alpha\rho}\pi^{\rho\beta}
\ee
It is straightforward to calculate the Poisson Brackets of $E$
with $H_{0},~H_{\alpha}$:
\be
\begin{array}{l}
  \{E, H_{\alpha}\}=-\frac{1}{2}\lambda^{\beta}_{\alpha}H_{\beta} \\
  \{E, H_{0}\}=-2a\gamma R=-2a\gamma (q^{1}+2q^{2})
\end{array}
\ee

At this point, it is crucial to observe that we can construct a
classical integral of motion, i.e. an extra gauge symmetry of the
corresponding classical action: notice that the trace of the
canonical momenta, $\gamma_{\mu\nu}\pi^{\mu\nu}$, has vanishing PB
with $H_{\alpha},~E$, and a --similar to $E$-- PB with $H_{0}$
equal to $2\gamma R$; thus, if we define:
\be
T=E-a\gamma_{\mu\nu}\pi^{\mu\nu}
\ee
we can easily derive the following PB of $T$ with
$H_{0},~H_{\alpha}$:
\be
\begin{array}{l}
  \{T, H_{0}\}=0 \\
  \{T, H_{1}\}=-\frac{a}{2}H_{1}\approx 0 \\
  \{T, H_{2}\}=-\frac{a}{2}H_{2}\approx 0\\
  \{T, H_{3}\}=0
\end{array}
\ee

The quantity $T$, is thus revealed to be first-class, and
therefore an integral of motion (since the Hamiltonian, is a
linear combination of the constraints):
\be
\dot{T}=\{T,H\}\approx 0 \Rightarrow T=\text{const}=C_{T}
\ee
The quantum version of $T$, is taken to be \cite{hajiceck}:
\be
\widehat{T}=\epsilon^{\alpha}_{\beta}\gamma_{\alpha\rho}\frac{\partial}{\partial
\gamma_{\beta\rho}}-a\gamma_{\alpha\beta}\frac{\partial}{\partial
\gamma_{\alpha\beta}}
\ee
(without any loss of generality --see (\ref{17paper3})-- we can safely
suppress $a$, whenever convenient, by setting it to 1).

Following the spirit of Dirac, we require:
\be
\widehat{T}\Psi=\epsilon^{\alpha}_{\beta}\gamma_{\alpha\rho}\frac{\partial
\Psi}{\partial
\gamma_{\beta\rho}}-a\gamma_{\alpha\beta}\frac{\partial
\Psi}{\partial
\gamma_{\alpha\beta}}=(q^{1}\frac{\partial\Psi}{\partial
q^{1}}+q^{2}\frac{\partial\Psi}{\partial
q^{2}}-\gamma\frac{\partial\Psi}{\partial \gamma})=C_{T}\Psi
\ee
The general solution \cite{carabedian} to the above equation has the form:
\be \label{25paper3}
\Psi=\gamma^{-C_{T}}\Theta(\frac{q^{1}}{q^{2}},\gamma q^{2})
\ee
$\Theta$ being an arbitrary function in its arguments.

Now, the number of our dynamical variables, is reduced from 3
($q^{i}, i=1,2,3$) to 2, namely the combinations
$w^{1}=q^{1}/q^{2}$ and $w^{2}=q^{2}q^{3}$. So, we have a further
reduction of the 3-dimensional configuration space spanned by the
3 $q$~'s. Again, in terms of the $w$~'s, the finally reduced
`'physical`' --although singular-- metric, is given by the
following relation:
\be \label{26paper3}
s^{kl}=g^{ij}\frac{\partial w^{k}}{\partial q^{i}}\frac{\partial
w^{l}}{\partial q^{j}}=\left(\begin{array}{cc}
  -16+4(w^{1})^{2} & 0 \\
  0 & 0
\end{array}\right)
\ee
The singular character of this metric is not unexpected; its
origin lies in the fact that
$L^{\alpha\beta\gamma\delta}T_{\alpha\beta}T_{\gamma\delta}=0$,
where
$L^{\alpha\beta\gamma\delta}=\frac{1}{4}(\gamma^{\alpha\gamma}\gamma^{\beta\delta}+
\gamma^{\alpha\delta}\gamma^{\beta\gamma}-2\gamma^{\alpha\beta}\gamma^{\gamma\delta})$
is the covariant supermetric (inverse to
$L_{\alpha\beta\gamma\delta}$) and $T_{\alpha\beta}$ are the
components of $T$, seen as vector field in the initial superspace
spanned by $\gamma_{\alpha\beta}$. Indeed, it is known that
reducing to null surfaces entails all sort of peculiarities.

So far, the degrees of freedom are two ($w^{1},~w^{2}$). The
vanishing of the $s^{22}$ component indicates that $w^{2}$ is not
dynamical at the quantum level. This fact has its analogue at the
classical level; indeed, consider the quantity:
\be \label{27paper3}
\Omega=\frac{(\gamma q^{2})^{\cdot}}{\gamma
q^{2}}=L_{\alpha\beta\mu\nu}\gamma^{\mu\nu}\pi^{\alpha\beta}-L_{\kappa\lambda\rho\sigma}
\pi^{\rho\sigma}\frac{C^{\alpha}_{\beta\mu}C^{\beta}_{\alpha\nu}\gamma^{\mu\kappa}
\gamma^{\nu\lambda}}{q^{2}}
\ee

In (\ref{27paper3}), the transition from velocity phase space to momentum
phase space has been made using the Hamiltonian (\ref{6paper3}) --with the proper $R$. It is
straightforward to verify that:
\be
\Omega=\frac{2}{a}T-\frac{4\varepsilon
\gamma^{23}}{q^{2}}H_{1}-\frac{4\varepsilon
\gamma^{13}}{q^{2}}H_{2}
\ee
Taking into account the weak vanishing of the linear constraints,
it is deduced that;
\be \label{29paper3}
\Omega=\frac{2}{a}T=\frac{2C_{T}}{a}=2C_{T}
\ee
if we set $a=1$.\\
Another way to show that $\Omega$ is constant, is the following:
\be
\{\Omega, H\}=\{\frac{\{\gamma q^{2}, H\}}{\gamma q^{2}},
H\}=\frac{4\gamma}{(\gamma
q^{2})^{2}}(4\gamma_{11}H_{1}^{2}+4\gamma_{22}H_{2}^{2}-8\varepsilon\gamma_{12}H_{1}H_{2})
\ee
Again, the weak vanishing of the linear constraints, ensures that
$\Omega$ is constant.

Using (\ref{27paper3}), (\ref{29paper3}) and the action, we have:
\be \label{31paper3}
\gamma q^{2}=C_{1}exp\{2C_{T}\int\widetilde{N}(t)dt\}
\ee
Now, returning to the quantum domain, we observe that out of the
three arguments of the wave function given in (\ref{25paper3}), only
$q^{1}/q^{2}$ is G.C.T. invariant --in the sense previously
explained. This suggests that we must somehow eliminate $\gamma,
~\gamma q^{2}$. To this end, we adopt the value zero for the
classical constant $C_{T}$. This amounts to restricting to a
2-parameter subspace of the classical space of solutions, spanned
by the 3 essential constants ~\cite{tchrisaut}. This means that we base
our quantum theory on this subspace and decree the wave function,
to be applicable to all configurations (classical or not). The
benefit of such an action, is twofold: $\gamma^{-C_{T}}$ drops
out, while at the same time $w^{2}\equiv\gamma q^{2}$ is set equal
to the constant $C_{1}$ --see (\ref{31paper3}). These facts, along with the
obligation that no derivatives with respect to $w^{2}$, are to
enter the Wheeler-DeWitt equation --see (\ref{26paper3})--, allow us to arrive
at the following form for the wave function (\ref{25paper3}):
\be
\Psi=\Theta(\frac{q^{1}}{q^{2}},C_{1})
\ee
and of course:
\be \label{33paper3}
w^{2}=q^{2}q^{3}=\text{const}=C_{1}
\ee
Now, the final reduction of the configuration space is achieved.
Our dynamic variable is the ratio $q^{1}/q^{2}$, which is a
combination of the only curvature invariants existing in Class A
Bianchi Type VI and VII, and emerges as the only true quantum
degree of freedom.

Consequently, following the spirit of \cite{hajiceck}, we have to
construct the quantum analogue of $H_{0}$ as the conformal
Laplacian, based on the non-singular part of the `'physical`'
metric (\ref{26paper3}), i.e.:
\be
\widehat{H}_{0}=-\frac{1}{2}\nabla^{2}_{c}+w^{2}(2+w^{1})
\ee
where:
\be \label{35paper3}
\nabla^{2}_{c}=\nabla^{2}=\frac{1}{\sqrt{s_{11}}}~\partial_{w^{1}}
\{\sqrt{s_{11}}~s^{11}~\partial_{w^{1}}\}
\ee
is the 1-dimensional Laplacian based on $s_{11}$
($s^{11}s_{11}=1$). Note that in 1-dimension the conformal group
is totally contained in the G.C.T. group, in the sense that any
conformal transformation of the metric can not produce any change
in the --trivial-- geometry and is thus reachable by some G.C.T.
Therefore, no extra term in needed in (\ref{35paper3}), as it can also
formally be seen by taking the limit $d=1,~R=0$ in the general
definition:
\be
\nabla^{2}_{c}\equiv\nabla^{2}+\frac{(d-2)}{4(d-1)}R=\nabla^{2}
\ee

Thus, the Wheeler-DeWitt equation, reads:
\be \label{36paper3}
\widehat{H}_{0}\Psi=\sqrt{(w^{2}-4)}~
\partial_{w}\{\sqrt{(w^{2}-4)}~\partial_{w}\}\Psi-\frac{C_{1}(2+w)}{2}\Psi=0
\ee
where, for simplicity, we set $w^{1}\equiv w$ and $w^{2}=C_{1}$.\\
Using the transformation $w=2cosh(z)$, the previous equation takes
the form:
\be \label{37paper3}
\frac{\partial^{2}\Psi}{\partial z^{2}}-(C_{1}+cosh(z))\Psi=0
\ee
The solutions to this family of equations are the Mathieu Modified
Functions --see \cite{abramowitz} and references therein, for an extended
treatise, in various cases.

As for the measure, it is commonly accepted that there is not a
unique solution. A natural choice, is to adopt the measure that
makes the operator in (\ref{36paper3}) hermitian, that is
$\mu(w)=\frac{1}{\sqrt{w^{2}-4}}$, or in the variable $z$,
$\mu(z)=1$. However, the solutions to (\ref{37paper3}) can be seen to
violently diverge for various values of $z \in [0,\infty)$, which
is the classically allowed region. If we wish to avoid this
difficulty, we can abandon hermiticity, especially in view of the
fact that we are interested in the zero eigenvalues of the
operator, and thus does not make any harm to lose realness of the
eigenvalues. If we adopt this attitude, we can find suitable
measures, e.g. $\mu(z)=e^{-z^{2}}$. The probability density
$\rho(z)=\mu(z)|\Psi(z)|^{2}$ is now finite, enabling one to
assign a number between $0$ and $1$ to
each Class A Type VI and VII geometry.\\
Another feature of the reduction to (\ref{37paper3}) is that the
final dynamical argument of $\Psi$ is the ratio $q^{1}/q^{2}$,
which is of degree zero in the scale factors, as seen from
equation (\ref{12paper3}). This consists a kind of build-in
regularization with respect to the volume of the 3-space.
Moreover, the solutions to this equation exhibit an increasingly
oscillatory behavior, as $C_{1}$ increases. This is most welcomed
and expected in view of $C_{1}$ being some kind of measure of the
3-volume, since it contains $\gamma$ --see (\ref{33paper3}).

\end{document}